\documentclass[twocolumn,showpacs,preprintnumbers,amsmath,amssymb]{revtex4}

\usepackage[dvips]{graphicx}
\usepackage{dcolumn}
\usepackage{bm}

\begin{document}
\def\bfB{\mbox{\bf B}}
\def\bfQ{\mbox{\bf Q}}
\def\bfD{\mbox{\bf D}}
\def\etal{\mbox{\it et al}}

\title{A simple mechanism for the reversals of Earth's magnetic field}

\author{Fran\c{c}ois P\'etr\'elis$^{1}$, St\'ephan Fauve$^{1}$, Emmanuel Dormy$^{2,3}$, Jean-Pierre Valet$^{3}$}

\address{(1)  Ecole Normale
Sup\'erieure, LPS, UMR CNRS 8550, 24 Rue Lhomond, 75005 Paris, France;
\\ (2) Ecole Normale Sup\'erieure, LRA, 24 rue Lhomond, 75005, Paris, France.
\\(3) Institut de Physique du Globe de Paris, UMR CNRS 7154, 4, Place
  Jussieu, F-75252 Paris Cedex 05, France.}

\date{\today}

\begin{abstract}
We show that a model, recently used to describe all the dynamical regimes of the magnetic field generated by the dynamo effect in the VKS experiment\cite{petrelis08}, also provides a simple explanation of the reversals of Earth's magnetic field, despite strong differences between both systems. 
\end{abstract}


\maketitle

The Earth's magnetic field can be roughly described as a strong axial dipole when averaged on a few thousands years. As shown by paleomagnetic records, it has frequently reversed its polarity on geological time scales. Field reversals have also been reported in
several numerical simulations of the geodynamo~\cite{roberts} and  more recently, in a laboratory experiment involving a von Karman swirling flow of
liquid sodium (VKS \cite{Berhanu}). It is worth pointing out that numerical simulations are performed in a parameter range orders of magnitude away from realistic values, and that both the parameter range and the symmetries of the flow in the VKS experiment strongly differ from the ones of the Earth's core. We thus expect that if a general mechanism for field reversals exists, it should not depend on details of the velocity field. This is expected in the vicinity of the dynamo threshold where nonlinear equations govern the amplitudes of the unstable magnetic modes. We assume that two modes have comparable thresholds. This has been observed for dipolar and quadrupolar dynamo modes~\cite{roberts72} and has been used to model the dynamics of the magnetic fields of the Earth or the Sun~\cite{dipquad}. However, in contrast to these previous models, we consider two axisymmetric stationary modes and expand the magnetic field $\bfB(r, t)$ as
\begin{equation}
\bfB(r, t) = a(t) {\bf B}_1(r) + b(t) {\bf B}_2(r) + \dots \,.
\end{equation}  
We define $A(t) = a + i b$ and write the evolution equation for $A$ using the symmetry constraint provided by the invariance $\bf B \rightarrow - \bf B$ of the equations of magnetohydrodynamics. This imposes $A \rightarrow -A$, thus the amplitude equation for $A$ is to leading nonlinear order
\begin{equation}
\dot{A} = \mu A+\nu \bar{A}+\gamma_1 A^3+\gamma_2 A^2\bar{A}+\gamma_3 A \bar{A}^2+\gamma_4\bar{A}^3\,
\label{amplitude}
\end{equation}
where $\mu$, $\nu$ and $\gamma_i$ are complex coefficients.   
Equations of the form (\ref{amplitude}) arise in different contexts, for instance for strong resonances, and their bifurcation diagrams are well documented~\cite{arnold}. Defining $A = R \exp i \theta$, 
a further simplification can be made when the amplitude $R$ has a shorter time scale than the phase $\theta$ and can be adiabatically eliminated. In that case, $\theta$ obeys an equation of the form
\begin{equation}
\dot \theta = \alpha_0 + \sum_{n\geq 1} \left(\alpha_n \cos 2n\theta + \beta_n \sin 2n\theta \right). 
\label{phase}
\end{equation}
The absence of odd Fourier terms results from the invariance $\bf B \rightarrow - \bf B$ that implies $\theta \rightarrow \theta + \pi$. Stationary solutions of (\ref{phase}) disappear by saddle-node bifurcations when parameters are varied. When no stationary solution exist any more, a limit cycle is generated which connects the former stable point $\theta_s$ to $\theta_s + \pi$, i.e., $B_s$ to $- B_s$ (see Fig. 1). This elementary mechanism for reversals is not restricted to the validity of (\ref{phase}) but results from the two dimensional phase space of (\ref{amplitude}) \cite{petrelis08}.
Thus, the qualitative features of the dynamics can be captured using the simplest possible model keeping the leading order Fourier coefficients $\alpha_0$ and $\beta_1$ ($\alpha_1$ can be eliminated by changing the origin $\theta \rightarrow \theta + \theta_0$). 

\begin{figure}
\centerline{a.\includegraphics[width=3.5cm]{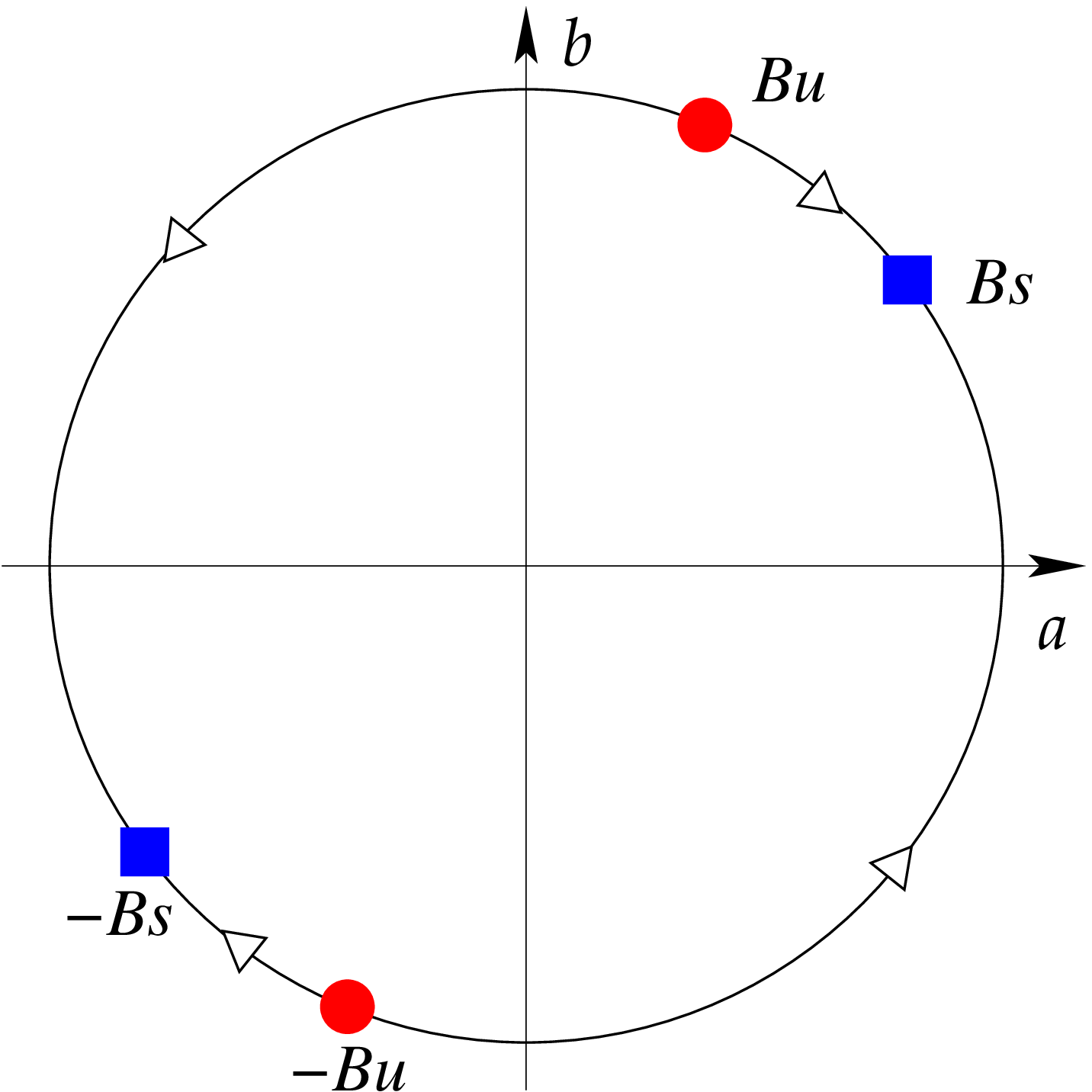}
\ \ \ \ 
b.\includegraphics[width=3.5cm]{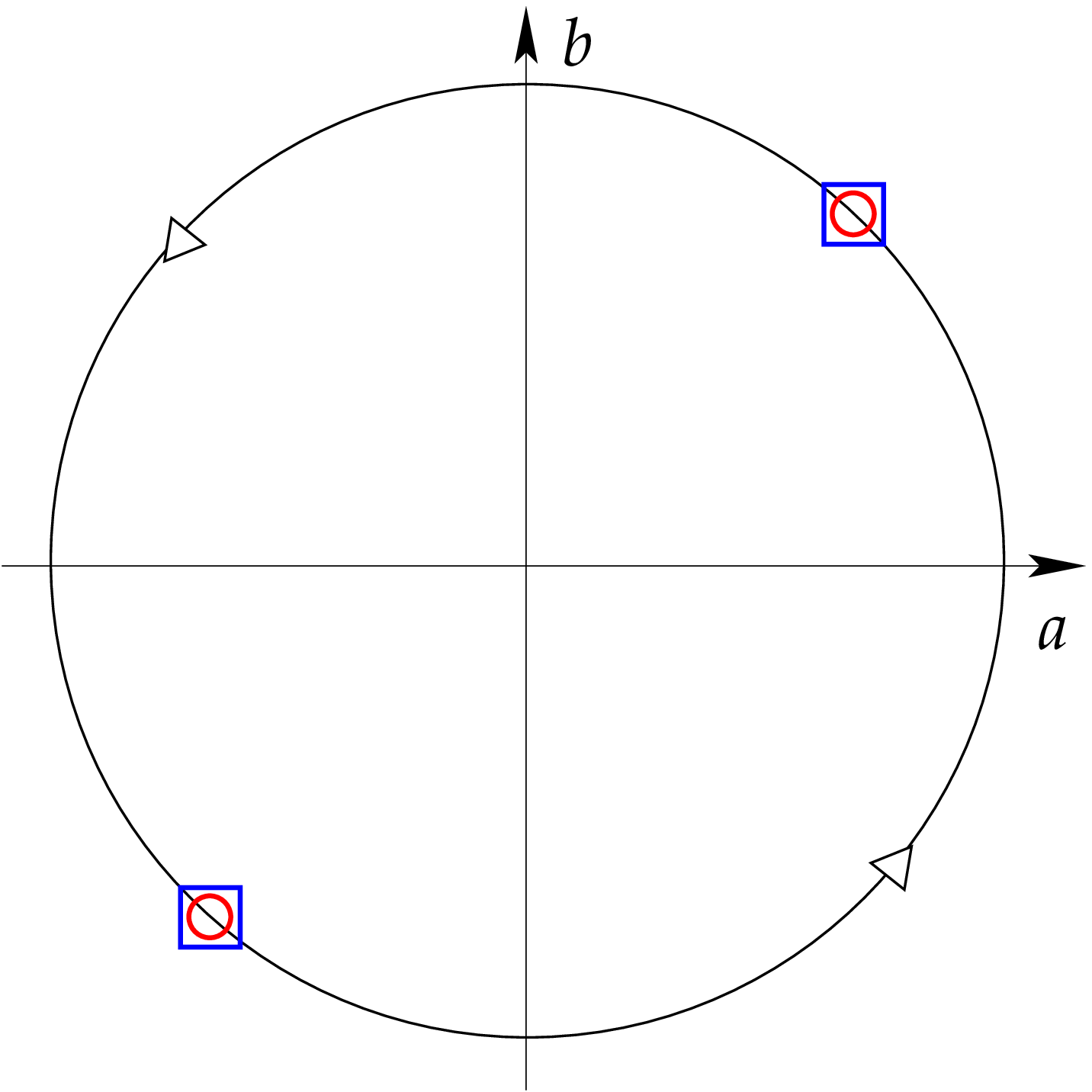}
}

\caption{Phase space of a system invariant under $\bfB\rightarrow -\bfB$
 and displaying a
  saddle-node bifurcation: (a) below the onset of the saddle-node
  bifurcation, (square blue): stable fixed points; (red circle): unstable
  fixed points.  (b) Above the threshold of the bifurcation, the fixed points
  (empty symbols) have collided and disappeared, the solution describes a
  limit cycle. Note that in (a), below the onset of the saddle-node
  bifurcation, fluctuations can drive the system from $Bs$ to $Bu$ (phase
  $Bs\rightarrow Bu$) and initiate a reversal (phase $Bu\rightarrow -Bs$) or an excursion (phase
  $Bu\rightarrow Bs$).} 
\end{figure}

So far, we did not consider possible effects of fluctuations. The flow in
the Earth's core, as well as in the VKS experiment, is far from being
laminar. We can therefore assume that turbulent fluctuations act as noisy
terms in the low dimensional system that describes the coupling between the
two magnetic modes. In Fig. 1a, the system is below the threshold of the
saddle-node bifurcation and in the absence of fluctuation exhibits two
stable (mixed) solutions. If the solution is initially located close to one of
the stable fixed points, say $Bs$, fluctuations can push the system away from
$Bs$. If it goes beyond the unstable fixed point $Bu$, it is attracted by the
opposite fixed point $-Bs$, and thus achieves a polarity reversal. A reversal
is made of two successive phases. The first phase $Bs\rightarrow Bu$ in Fig. 1a is the
approach toward an unstable fixed point. The deterministic dynamics acts
against the evolution and this phase is slow. The second phase
$Bu\rightarrow - Bs$, is fast since the deterministic dynamics favors the motion.

At the end of the first phase, the system may
return toward the initial stable fixed point (phase $Bu\rightarrow Bs$), which corresponds to an
excursion.   We emphasize that, close enough to the saddle-node bifurcation, reversals require vanishingly small fluctuations. 
To take them into account, we modify the equation for $\theta$
into
\begin{equation}
\dot \theta = \alpha_0 + \alpha_1 \sin(2 \theta) + \Delta \zeta (t) \, ,
\end{equation} 
from which we derive the evolution of the dipole
by $d=R \cos(\theta + \theta_0)$. $\zeta$ is a Gaussian white noise and
$\Delta$ is its amplitude. We
have computed a time series of the dipole amplitude  for a
system below the threshold of the bifurcation ($\alpha_1=-185$ Myr$^{-1}$, $\alpha_0 / \alpha_1= - 0.9, \,   \, \theta_0=0.3)$ and with noise amplitude $\Delta / \sqrt{\vert \alpha_1 \vert}=0.2.$ Note that $\alpha_1$ is arbitrary at this stage. Its value results from a fit of paleomagnetic data (see below). The dipole amplitude is displayed in Fig. 2 together with a time series of the magnetic field measured in the VKS experiment and the composite record of the geomagnetic dipole for the past 2 Myr. The three curves
display very similar behaviors with abrupt reversals and large
fluctuations.  We have checked that similar dynamics are obtained when equation (\ref{amplitude}) with noisy coefficients is numerically integrated.

\begin{figure}
\centerline{\includegraphics[width=6.5cm]{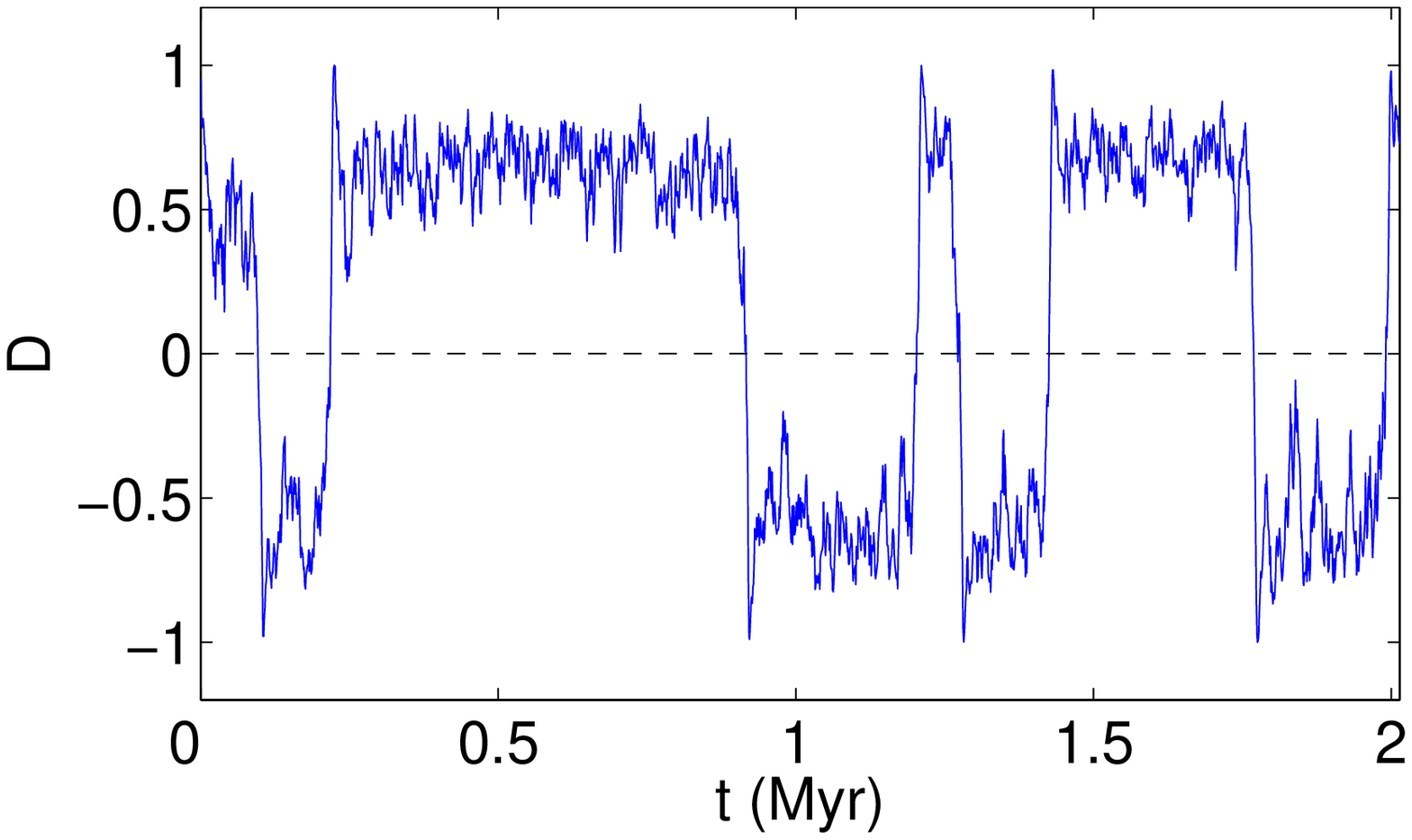}}
\centerline{\includegraphics[width=6.5cm]{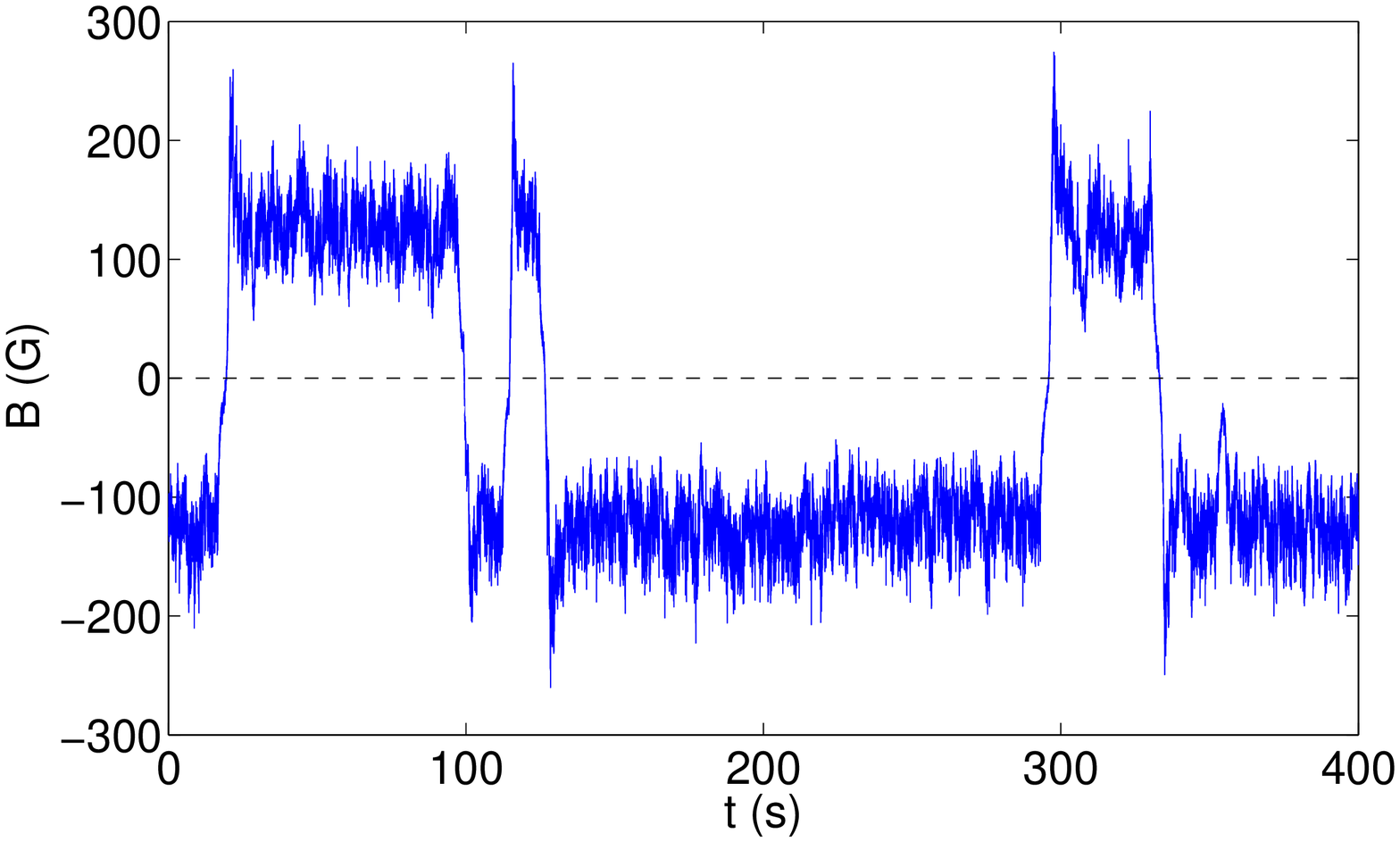}}
\centerline{\includegraphics[width=6.5cm]{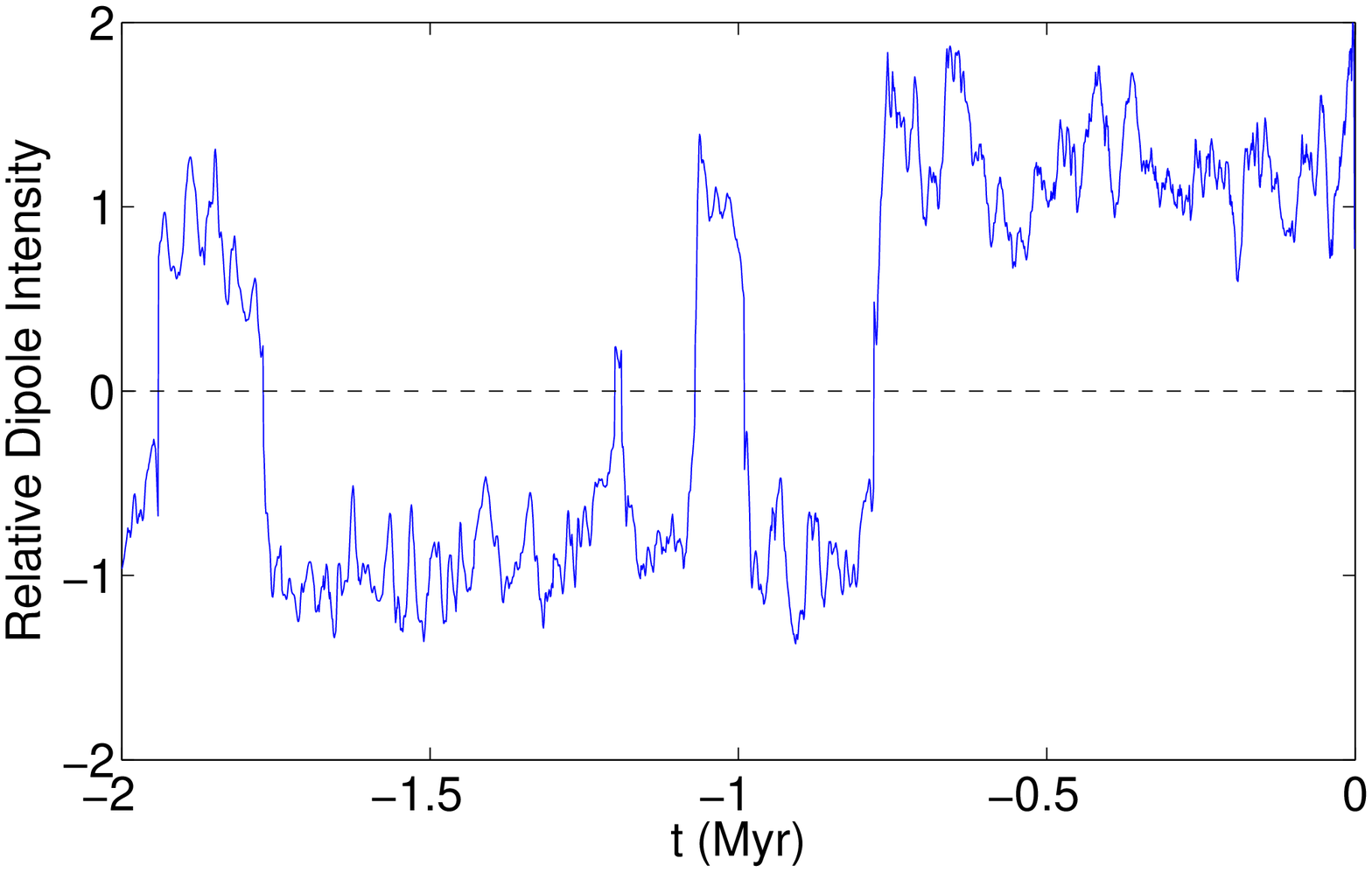}}
\caption{(Top) Time series of the dipole amplitude for a system below
the threshold of the bifurcation (see text for the values of the parameters) (Middle) Time series of the magnetic field measured in the VKS experiment for the two impellers rotating with different frequencies F1=22 Hz, F2=16 Hz  (data from \cite{Berhanu}). (Bottom) Composite paleointensity curve for the past 2 millions years, present corresponds to t=0 (data from \cite{Valet2005}).} 
\end{figure}

One of the most noticeable features common to these three curves is the existence
of a significant overshoot that immediately follows the reversals. In
Fig. 3, the enlarged views of the period surrounding reversals and
excursions also show that this is not the case for the excursions. In fact,
the relative position of the stable and unstable fixed points (Fig. 1)
controls the evolution of the field. During the first phase, reversals and
excursions are similar, but they differ during the second phase. The
synopsis shows that the reversals reach the opposite fixed point from a
larger value and thus display an overshoot while excursions do not.  

\begin{figure*}[!htb]
\centerline
{\includegraphics[width=5.5cm]{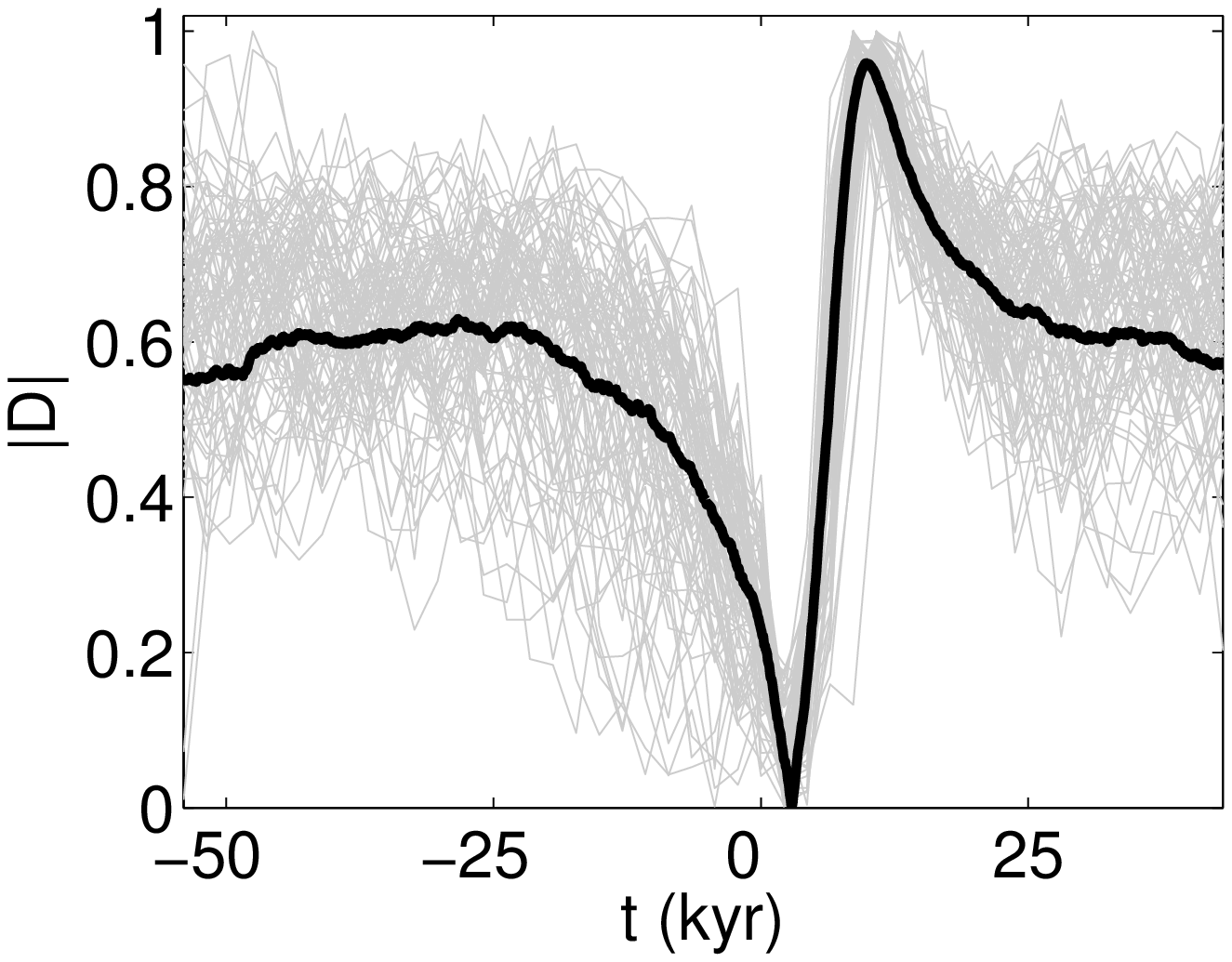}\ \includegraphics[width=5.5cm]{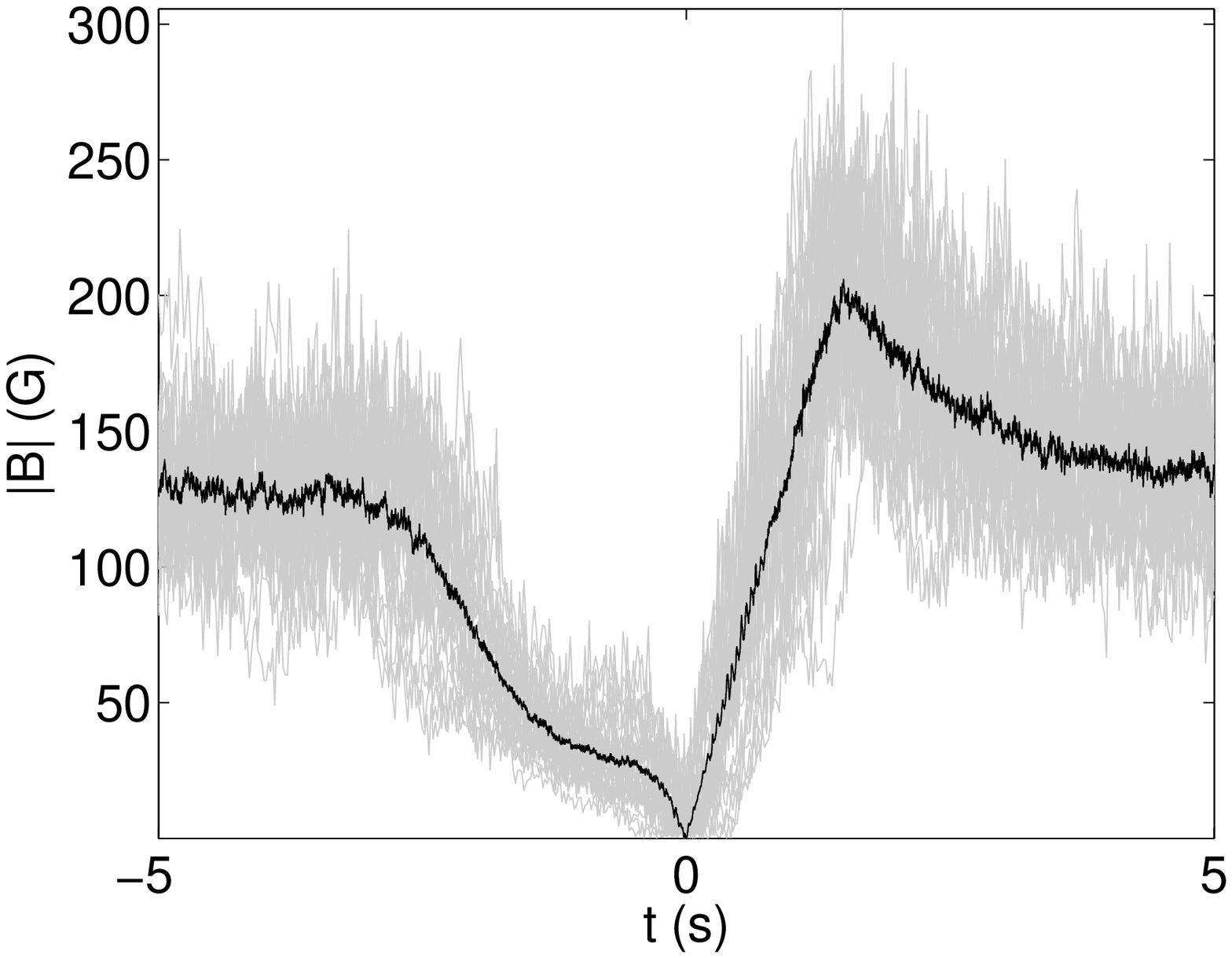}\
\includegraphics[width=5.5cm]{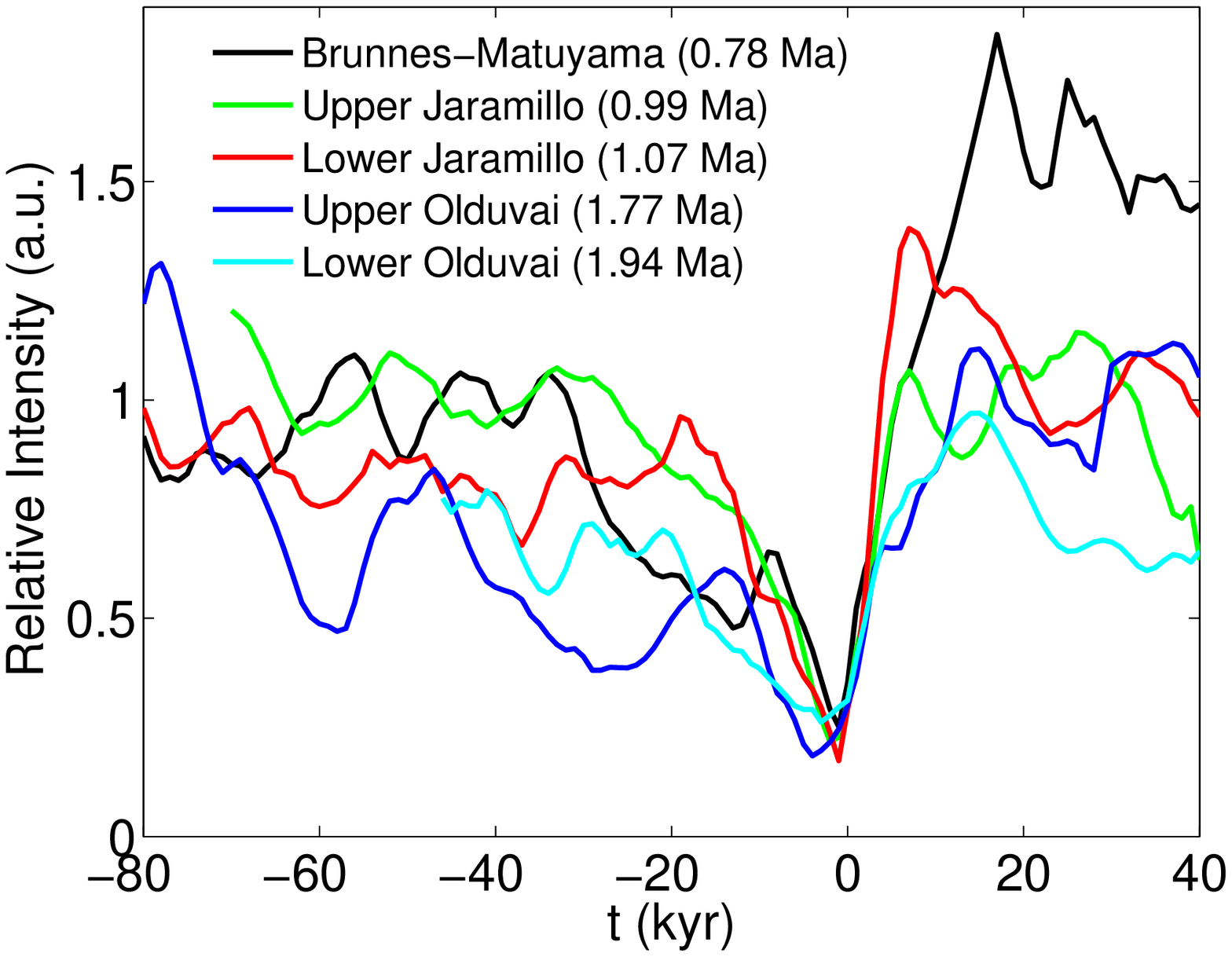}}
\centerline
{\includegraphics[width=5.5cm]{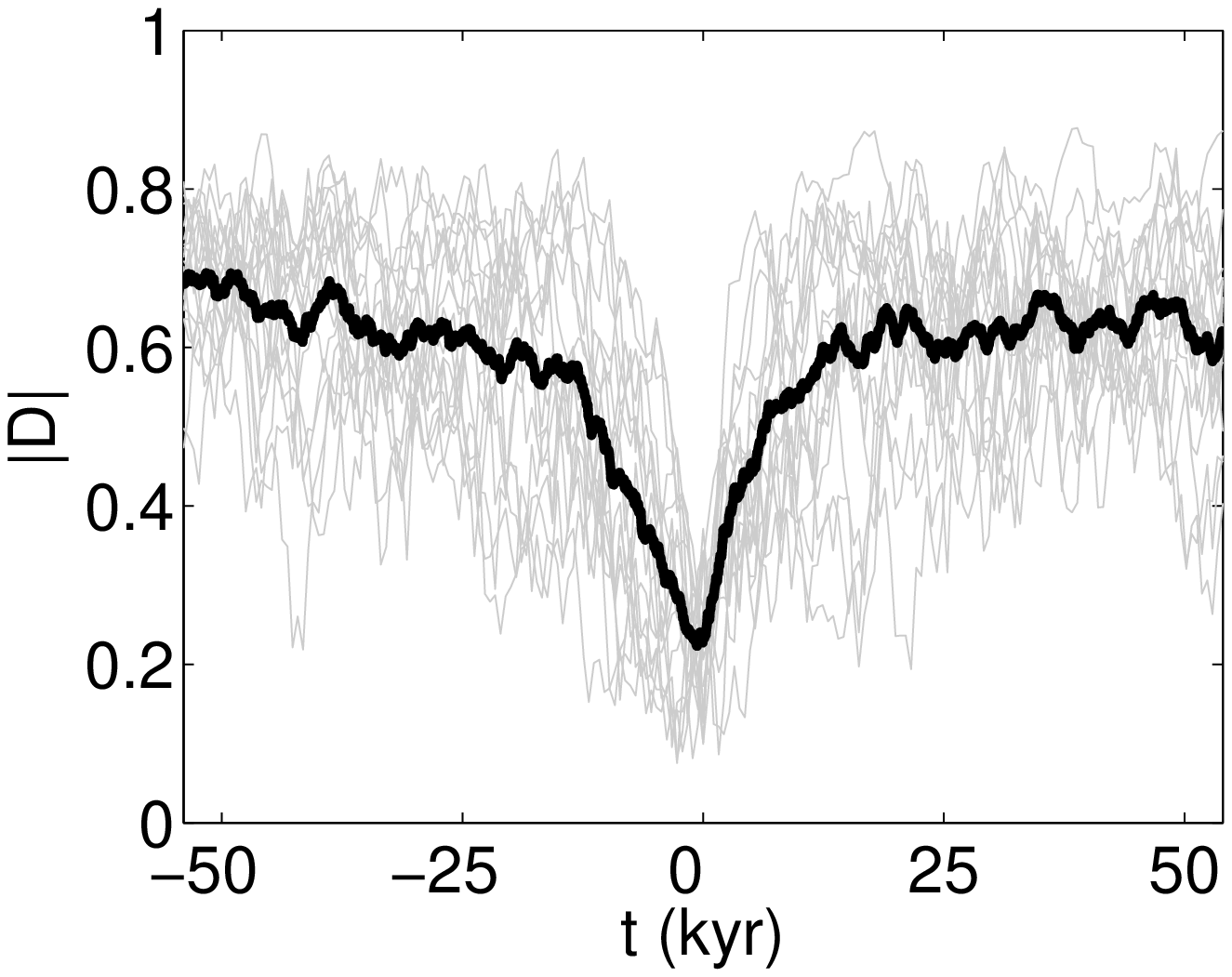}\ \includegraphics[width=5.5cm]{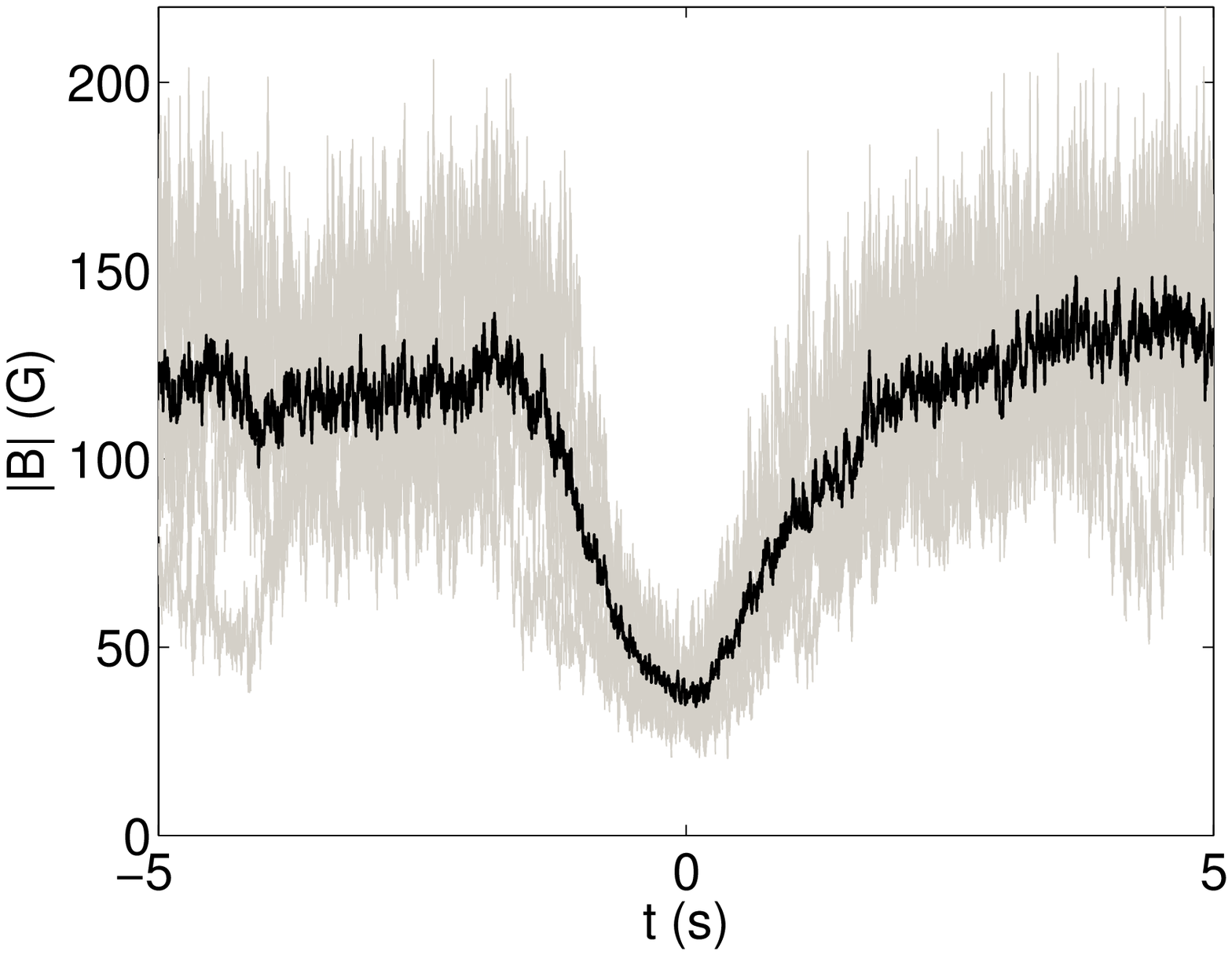}\
\includegraphics[width=5.5cm]{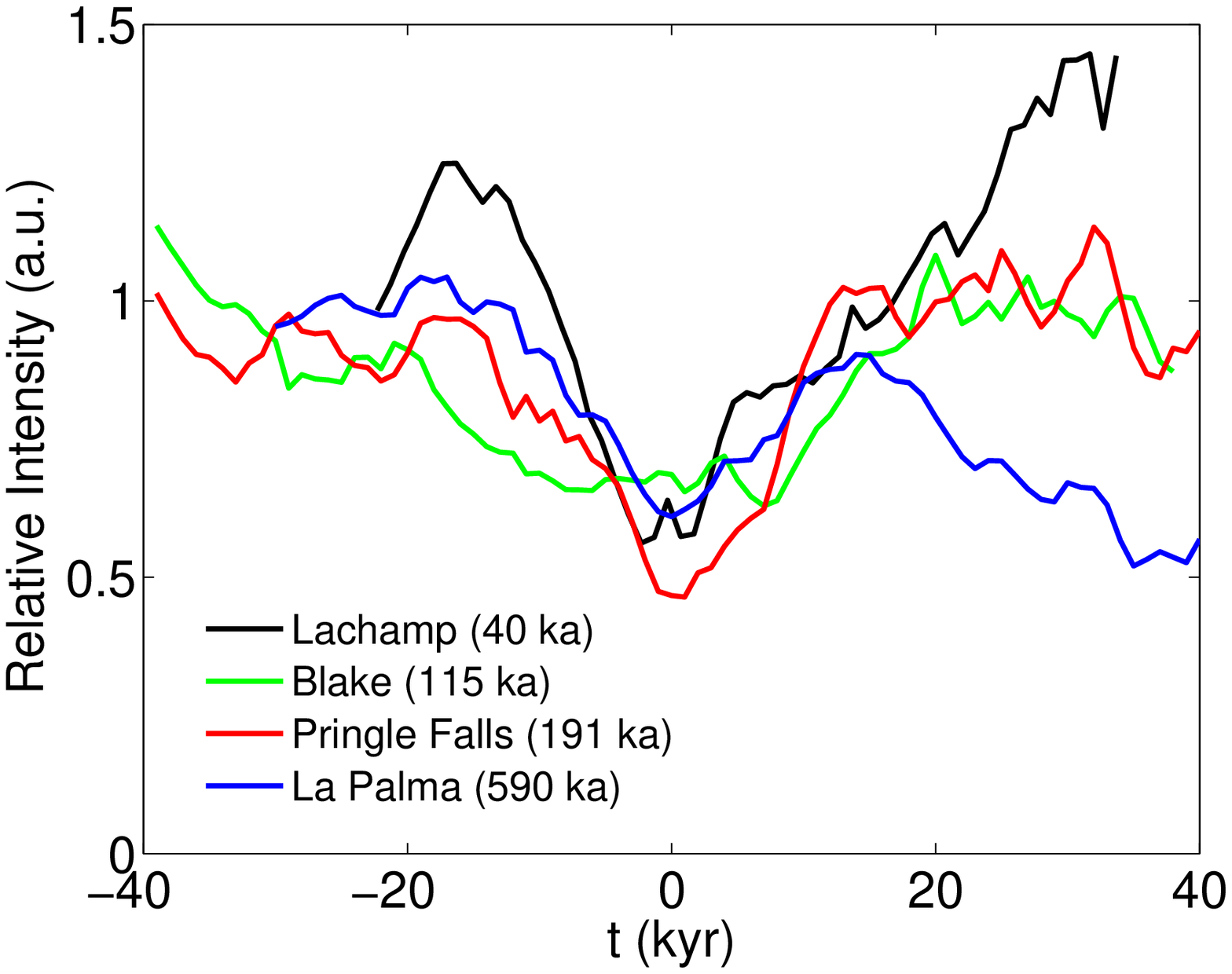}}
\caption{\label{fig:wide}Comparison of reversals and excursions in the solutions of the equation for $\theta$ (left), the VKS experiment (middle) (data from \cite{Berhanu}), and in paleomagnetic data (right) (data from \cite{Valet2005}). Black curves represent the averaged curve, each realization being represented in grey.}
\end{figure*}

\begin{figure}[!htb]
\centerline{\includegraphics[width=7cm]{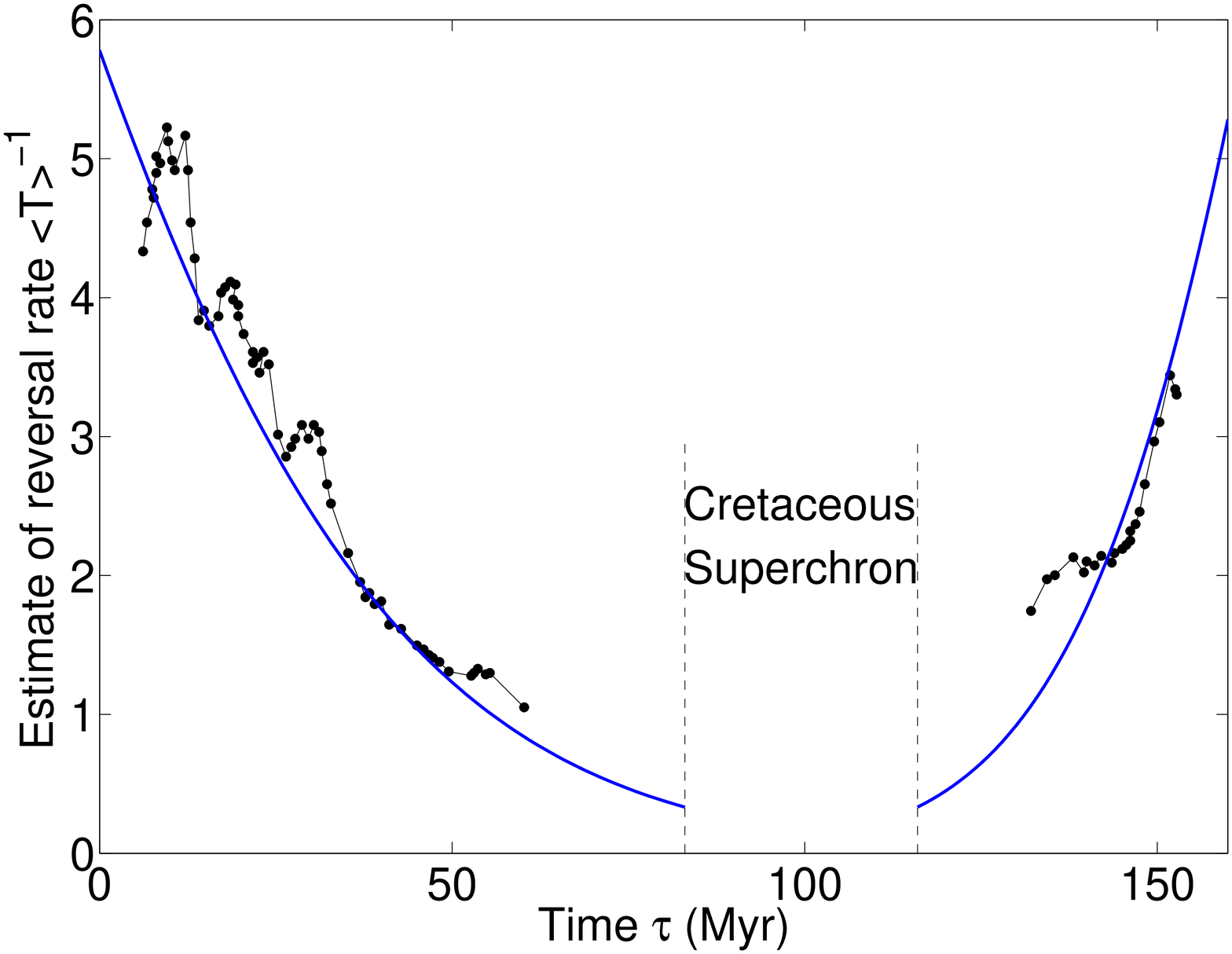}
}

\caption{Variation of the reversal rate $\langle T\rangle^{-1}$ close to a superchron together with the fit with equation (\ref{tempsmoyen}). Data ($\bullet$) have been extracted from \cite{mcfadden}. $\vert \alpha_1\vert = 185$ Myr$^{-1}$, $\Delta / \sqrt{\vert \alpha_1 \vert}=0.2$, $\alpha_0 / \alpha_1 = -0.9 (1 - \tau / 650)$ for $\tau<80$, 
$\alpha_0 / \alpha_1 = -0.9 (0.55 + \tau / 360)$ for $\tau>120$, where $\tau$ is time in millions of years.}
\label{superchrons}
\end{figure}

Below the onset of bifurcation, reversals 
occur very seldomly, which indicates that their occurence requires rarely
cooperative fluctuations. The evolution from the stable to the
unstable fixed point (phase $Bs\rightarrow Bu$ in Fig. 1) can be described as the noise driven
escape of the system from a metastable potential well. The durations of
polarity intervals are equivalent to the exit time and are exponentially
distributed \cite{vankampen} according to 
\begin{equation}
P[T] \propto {\exp({-T}}/{\langle T \rangle})\,.
\end{equation} 
The averaged duration $\langle T\rangle$ depends on the
intensity of the fluctuations and on the 
distance to the saddle-node bifurcation. For the model studied, we obtain
\begin{equation}
\langle T\rangle = \frac{\pi}{\sqrt{\alpha_1^2-\alpha_0^2}}\, \exp \left[2 \vert \alpha_1\vert \frac{(2(\alpha_1+\alpha_0)/\alpha_1)^{3/2}}{3\Delta^2}\right] \, ,
\label{tempsmoyen}
\end{equation} 
which corresponds to $\langle T \rangle \simeq 170$ kyr for the parameters used in Fig. 2.
We observe that deterministic parameters are of the order of the Ohmic dissipation time ($\pi / \vert \alpha_1 \vert \simeq 17000$ years) whereas much larger time scales are measured for $\langle T \rangle$ because of the low noise intensity. This explains that the mean duration of phases with given polarity is much larger than the one of a reversal. The above predictions assume that the noise
intensity and the deterministic dynamics do not vary in time. It is likely
that the Rayleigh number in the core and the efficiency of coupling
processes between the magnetic modes have evolved throughout the Earth's history. The exponential dependence of the mean polarity duration $\langle T\rangle$ on
noise intensity implies that a moderate change in convection can result in
a very large change of $\langle T\rangle$. This might account for changes in the rate of
geomagnetic reversals and very long periods without reversals (so called
superchrons). The reversal rate reported in \cite{mcfadden} is displayed in Fig. \ref{superchrons} together with a fit using Eq. \ref{tempsmoyen} assuming a linear variation in time of the coefficients governing the distance between $Bs$ and $Bu$.  A simple variation of one parameter captures the temporal evolution of the reversal rate. Thus,  although it can be claimed that there are several fitting parameters, this quantitative agreement strengthens the validity of our model of reversals.

We now consider the different possibilities for the mode $\bf{B}_2(r)$ coupled to the Earth's dipolar field $\bf{B}_1(r)$.
Assuming that the equator is a plane of mirror symmetry, the different modes can be classified as
follows: dipolar modes are the ones unchanged by mirror symmetry, $\bfD \rightarrow \bfD$, whereas quadrupolar modes change sign, $\bfQ \rightarrow -\bfQ$. From an analysis of paleomagnetic data, McFadden et al. have proposed that reversals involve an interaction between dipolar and quadrupolar modes~\cite{merrill}. In that case,  $\bf{B}_1(r)$ and $\bf{B}_2(r)$ change differently by mirror symmetry. If the flow is mirror symmetric, this implies that Eq. (\ref{amplitude}) should be invariant under $A \rightarrow \bar A$ which amounts to $\theta \rightarrow - \theta$. Consequently, $\alpha_n = 0$ in (\ref{phase}) and no limit cycle can be generated. 
We thus obtain an interesting prediction in that case: if reversals involve a coupling of the Earth's dipole with a quadrupolar mode, then this requires that the flow in the core has broken mirror symmetry. 
This mechanism explains several observations made in numerical simulations: reversals of the axial dipole, simultaneous with the increase of the axial quadrupole, have been found when the North-South symmetry of the convective flow is broken~\cite{Li}. It has been shown that if the flow or 
the magnetic field is forced to remain equatorially symmetric, then reversals do not occur \cite{Nishikawa}.  The possible effects of heterogeneous heat flux at the core-mantle boundary (CMB) on the dynamics of the Earth's magnetic field have also been investigated numerically \cite{Glatzmaier99}. Compared to the homogeneous heat flux, patterns of antisymmetric heterogeneous heat flux were shown to yield more frequent reversals. Within our description, this appears as a direct consequence of the breaking of hydrodynamic equatorial symmetry driven by the thermal boundary conditions. 
From the point of view of the observations, little is known on the actual flow inside the Earth's core. It has recently been noted that the ends of superchrons are followed by major flood basalt eruptions and massive faunal depletions \cite{Courtillot07}. The authors suggested that large thermal plumes ascending through the mantle favor reversals and subsequently produce large eruptions. In the light of our work, it is tempting to associate to the thermal plumes (which provide a localized thermal forcing at the surface of the CMB) with an enhanced deviation from the flow equatorial symmetry, which results according to the above description in an increase of reversal frequency and therefore ends superchrons.

In contrast, another scenario has been proposed in which the Earth's dipole is coupled to an octupole, i.e., another mode with a dipolar symmetry \cite{clement}. This does not require additional constraint on the flow in the core in the framework of our model. In any case, the existence of two coupled modes allows the system to evolve along a path that avoids ${\bf B}={\bf 0}$. In physical space, this means that the total magnetic field does not vanish during a reversal but that its spatial structure changes. 

Numerical simulations of MHD equations~\cite{sarson} or of mean field models have displayed reversals that seem to involve ``transitions between the steady and the oscillatory branch of the same eigenmode"~\cite{stefani}. That situation can be obtained in the vicinity of a codimension-two bifurcation with a double zero eigenvalue and only one eigenmode. This type of bifurcation also exists in our model (\ref{amplitude}) but requires tuning of two parameters $\mu$ and $\nu$. This is not necessary for the scenario of reversals we have described. Other features of reversals observed in numerical simulations at magnetic Prandtl number of order one, such as mechanisms of advection/amplification of the field due to localized flow processes \cite{aubert}, are not described by our model which requires the limit of small magnetic Prandtl number (relevant to the Earth's core). 

Equations similar to (\ref{phase}) have been studied in a variety of problems, for instance for the orientation of a rigid rotator subject to a torque \cite{Hoyng}, used as a toy model for the toroidal and the poloidal field of a single dynamo mode. Indeed, symmetries constrain the form of the equation for $\theta$ even though the modes and the physics involved are different.
We emphasize that the above scenario is generic and not restricted to the equation considered here. 
Limit cycles generated by saddle-node bifurcations that result from the coupling between two modes occur in Rayleigh-B\'enard convection~\cite{convection}. A similar mechanism can explain reversals of the large scale flow generated over a turbulent background in thermal convection or in periodically driven flows~\cite{hydro}.
We have proposed a scenario for reversals of the magnetic field generated
by dynamo action that is based on the same type of bifurcation structure in the presence of noise.
It offers a simple and unified explanation for many intriguing features of the Earth's magnetic field. The most significant output is that the mechanism predicts specific characteristics of the field obtained from paleomagnetic records of reversals and from recent experimental results. Other characteristic features such as excursions as well as the existence of superchrons are understood in the same framework.  
Below the threshold of the saddle-node bifurcation, fluctuations drive random reversals by excitability. We also point out that above its threshold, the solution is roughly periodic. It is tempting to link this regime to the evolution of the large scale dipolar field of the Sun (which reverses polarity roughly every 11 years). Recent measurements of the Sun surface magnetic field have shown that two components oscillate in phase-quadrature \cite{stenflo}.  This would be coherent with the oscillatory regime above the onset of the saddle-node bifurcation if these components correspond to two different modes.\\

\vspace{-0.5cm}
We thank our colleagues from the VKS team with whom the data published in \cite{Berhanu} have been obtained.

\end{document}